\documentclass[twocolumn]{jpsj3}
\usepackage{txfonts}
\usepackage{braket}

\title{On the Puzzle of Odd-Frequency Superconductivity}

\def\runauthor{H. \textsc{Kusunose} {\it et al.}}
\author{Hiroaki \textsc{Kusunose}\thanks{kusu@phys.sci.ehime-u.ac.jp},
Yuki \textsc{Fuseya}$^{1}$ and
Kazumasa \textsc{Miyake}$^{1}$
}

\inst{Department of Physics, Ehime University, Matsuyama, Ehime 790-8577, Japan\\
$^{1}$Division of Materials Physics, Department of Materials Engineering Science,\\Graduate School of Engineering Science, Osaka University,\\Toyonaka, Osaka 560-8531, Japan
}

\abst{
Since the first theoretical proposal by Berezinskii, an odd-frequency superconductivity has encountered the fundamental problems on its thermodynamic stability and rigidity of a homogenous state accompanied by unphysical Meissner effect.
Recently, Solenov {\it et al}. [Phys. Rev. B {\bf 79} (2009) 132502.] have asserted that the path-integral formulation gets rid of the difficulties leading to a stable homogenous phase with an ordinary Meissner effect.
Here, we show that it is crucial to choose the appropriate saddle-point solution that minimizes the effective free energy, which was assumed {\it implicitly} in the work by Solenov and co-workers.
We exhibit the path-integral framework for the odd-frequency superconductivity with general type of pairings, including an argument on the retarded functions via the analytic continuation to the real axis.
}

\kword{odd-frequency superconductivity, Stratonovich-Hubbard transformation, saddle-point approximation, Meissner kernel}

\begin{document}
\maketitle

\section{Introduction}
The field of anisotropic superconductivity has developed extensively in condensed matter physics.
In the anisotropic superconductors, a strong Coulomb repulsion with relatively local character tends to suppress an on-site amplitude of the Cooper-pair wavefunction, and it favors an off-site pairing instead.
It is also possible to form an off-time or equivalently an odd-frequency pairing to avoid the strong repulsion.
It was first proposed by Berezinskii in the context of the superfluid $^{3}$He\cite{Berezinskii74}, and was revisited later for the mechanism of the cuprate superconductors by Balatsky and Abrahams\cite{Balatsky92}.
After the proposal of the odd-frequency superconductivity, considerable amount of investigations have been carried out both theoretically\cite{Emery92,Abrahams93,Coleman94,Abrahams95,Vojta99,Fuseya03,Shigeta09,Kalas08} and experimentally\cite{Kawasaki03} in a wide variety of theoretical models and realistic materials.
The anomalous proximity effect in the superconducting junctions and superfluid $^{3}$He has also been investigated by the odd-frequency pairing mechanism\cite{Higashitani09,Asano06,Tanaka07,Eschrig07,inhomo}.

In contrast to proceeding of such stimulating investigations, fundamental problems lying in the odd-frequency superconducting state remain unresolved.
The nuclei of the problems are thermodynamic instability of the homogeneous odd-frequency superconducting state\cite{Heid95,Heid95a,Coleman94} and unphysical Meissner effect with the negative Meissner kernel\cite{Abrahams95}.
In other words, these issues end up with a pure imaginary penetration depth and a deficiency of rigidity of a homogeneous order parameter, and that the free energy of the superconducting phase is lower than that of the normal phase in the high-temperature limit.
The common source of these problems is ascribed to the relation among the anomalous green functions,
\begin{equation}
F^{+}(i\omega_{n})=F(-i\omega_{n})^{*}=-F(i\omega_{n})^{*},
\label{increl}
\end{equation}
in the case of the odd-frequency pairing (see the discussions below).
In the case of the even-frequency pairing, the opposite sign appears in the most right-hand side, and no problems as mentioned in the above do occur.

Recently, Solenov {\it et al}. have demonstrated that the path-integral treatment of the odd-frequency superconductivity ends up with the different relation from (\ref{increl}), i.e.,
\begin{equation}
F^{+}(i\omega_{n})=F(i\omega_{n})^{*},
\end{equation}
which is the same relation for the even-frequency pairing\cite{Solenov09}.
As a result, a homogeneous odd-frequency superconductivity is thermodynamically stable in the bulk and exhibits the ordinary Meissner effect.
In their argument, the lack of an appropriate hamiltonian with broken U(1) gauge symmetry is vital in the case of the odd-frequency pairing, and only the path-integral approach provides reasonable results.
Although this is an important and correct statement, the essence of the problem still seems to remain obscure.

In this paper, we aim to unveil the essence of the problems by emphasizing the choice of the saddle-point solution for the odd-frequency pairing.
Although the correct choice between the gap function $\Delta(i\omega_{n})$ and its conjugate counterpart $\overline{\Delta}(i\omega_{n})$ is not apparent, Solenov {\it et al}. have {\it implicitly} assumed $\overline{\Delta}(i\omega_{n})=\Delta(i\omega_{n})^{*}$ in their argument [eq.~(10) in ref.~\citen{Solenov09}].
In view of the delicate issue to remedy the fate of the odd-frequency superconducting state, we should demonstrate a detailed formulation for the odd-frequency pairing state with great care in the path-integral framework\cite{Negele88,Nagaosa99}.
In the next section, we give the derivation of the free energy, and discuss the correct choice of the saddle-point solution for the case of an odd-frequency equal spin pairing as the simplest example.
With the correct choice of the saddle point, we show that the Meissner kernel becomes positive as similar to the even-frequency pairing.
The green functions and their analytic continuation to the real axis are also given.
We summarize the paper in \S3.
The formulation for the general type of pairings is given in Appendix.

\section{Path-integral framework for superconductivity}
In this section, we derive the minimal formula to argue the cure for the odd-frequency superconductivity.
In order to focus on the essence of the issue, we consider the case of the triplet pairing only with $\uparrow$ spin pairs, i.e., the pairing interaction $V$ ($<0$) is assumed to work only among $\uparrow$ spins (it is equivalent to the case of the spinless fermions).

\subsection{Free energy}

We first give the derivation of the effective free energy functional for superconductivity.
Let us begin with the partition function in the path-integral form,
\begin{align}
&Z=\int{\cal D}\overline{\psi}{\cal D}\psi\,e^{-(S_{0}+S_{\rm int})},
\cr
&\quad
S_{0}=\int_{1}\overline{\psi}_{\uparrow}(1)\left(\partial_{\tau}+\xi\right)\psi_{\uparrow}(1),
\cr
&\quad
S_{\rm int}=\frac{1}{2\beta}\int_{1}\int_{2}V(1-2)\,
\overline{\rho}_{\uparrow\uparrow}(1,2)\,\rho_{\uparrow\uparrow}(1,2),
\label{pathz}
\end{align}
where we have used the abbreviations for the space-time coordinate, $1\equiv x_{1}=(\mib{r}_{1},\tau_{1})$ and $\int_{1}\equiv\int_{0}^{\beta}d\tau_{1}\int d\mib{r}_{1}$ for notational simplicity.
$\beta=1/T$ is the inverse temperature, and the kinetic energy, $\xi=-\mib{\nabla}^{2}/2m-\mu$ is measured from the chemical potential, $\mu$.
The pairing interaction $V(1-2)$ is a real even function with respect to $(x_{1}-x_{2})$.
A strongly retarded interaction may be important to realize the odd-frequency pairing.
$\rho_{\uparrow\uparrow}(1,2)=\psi_{\uparrow}(1)\psi_{\uparrow}(2)$ and $\overline{\rho}_{\uparrow\uparrow}(1,2)=\overline{\psi}_{\uparrow}(2)\overline{\psi}_{\uparrow}(1)$ are the pair-density field and its complex-conjugate counterpart.
The anti commutation (Pauli principle) demands the relations, $\rho_{\uparrow\uparrow}(1,2)=-\rho_{\uparrow\uparrow}(2,1)$ and $\overline{\rho}_{\uparrow\uparrow}(1,2)=-\overline{\rho}_{\uparrow\uparrow}(2,1)$.

Using the Stratonovich-Hubbard transformation for the interaction part\cite{Negele88}, we have
\begin{align}
&e^{-S_{\rm int}}=\int{\cal D}\overline{\Delta}{\cal D}\Delta\,e^{-(S_{\rm aux}+S_{\Delta})},
\cr
&\quad
S_{\rm aux}
=-\frac{1}{2\beta}\int_{1}\int_{2}\left[\overline{\rho}_{\uparrow\uparrow}(1,2)\Delta_{\uparrow\uparrow}(1,2)+\overline{\Delta}_{\uparrow\uparrow}(1,2)\rho_{\uparrow\uparrow}(1,2)\right],
\cr
&\quad
S_{\Delta}
=-\frac{1}{2\beta}\int_{1}\int_{2}\frac{1}{V(1-2)}\overline{\Delta}_{\uparrow\uparrow}(1,2)\,\Delta_{\uparrow\uparrow}(1,2).
\label{pathaux}
\end{align}
Note that the auxiliary fields, $\Delta_{\uparrow\uparrow}(1,2)$ and $\overline{\Delta}_{\uparrow\uparrow}(1,2)$, are the complex $c$-number fields corresponding to the gap functions, and they satisfy the same symmetry relations of $\rho_{\uparrow\uparrow}(1,2)$ and $\overline{\rho}_{\uparrow\uparrow}(1,2)$.
At this point, we have introduced no approximations.

Now, let us find a homogenous saddle-point solution that is independent of the center-of-mass coordinate, $R\equiv(x_{1}+x_{2})/2$.
For this purpose, we replace the path integral over $\overline{\Delta}_{\uparrow\uparrow}(1,2)$ and $\Delta_{\uparrow\uparrow}(1,2)$ with trial path for the saddle point.
Then, we define the ``mean-field'' free-energy functional as
\begin{equation}
\beta{\cal F}_{\rm MF}[\overline{\Delta},\Delta]=-\ln Z_{\rm MF}=-\ln\left[\int{\cal D}\overline{\psi}{\cal D}\psi\,e^{-S_{\rm MF}(\overline{\psi},\psi,\overline{\Delta},\Delta)}\right].
\label{freeenergy}
\end{equation}
Here $S_{\rm MF}=S_{0}+S_{\Delta}+S_{\rm aux}$ is the ``mean-field'' action in which $\overline{\Delta}_{\uparrow\uparrow}(1,2)$ and $\Delta_{\uparrow\uparrow}(1,2)$ are replaced by the trial path, $\overline{\Delta}_{\uparrow\uparrow}(r)$ and $\Delta_{\uparrow\uparrow}(r)$, with the relative coordinate $r\equiv x_{1}-x_{2}$.
The true saddle point will be determined by minimizing the free-energy functional with respect to the trial path, $\overline{\Delta}_{\uparrow\uparrow}(r)$ and $\Delta_{\uparrow\uparrow}(r)$.

Let us introduce the fourier transformation,
\begin{align}
&\psi_{\uparrow}(1)=\frac{1}{\sqrt{\beta}}\sum_{k}\psi_{\uparrow}(k)e^{ikx_{1}},
\quad
\overline{\psi}_{\uparrow}(1)=\frac{1}{\sqrt{\beta}}\sum_{k}\overline{\psi}_{\uparrow}(k)e^{-ikx_{1}},
\cr
&\Delta_{\uparrow\uparrow}(r)=\sum_{k}\Delta_{\uparrow\uparrow}(k)e^{ikr},
\quad
\overline{\Delta}_{\uparrow\uparrow}(r)=\sum_{k}\overline{\Delta}_{\uparrow\uparrow}(k)e^{-ikr},
\label{ftsimple}
\end{align}
where $k=(\mib{k},i\omega_{n})$, $\omega_{n}=(2n+1)\pi/\beta$, $\sum_{k}=\sum_{\mib{k}}\sum_{n}$ and $kx=\mib{k}\cdot\mib{r}-\omega_{n}\tau$.
Then, the action containing the fermion fields is expressed as
\begin{equation}
S_{0}+S_{\rm aux}=
\frac{1}{2}\sum_{k}
\begin{pmatrix}
\overline{\psi}_{\uparrow}(k), & \psi_{\uparrow}(-k)
\end{pmatrix}
\hat{M}_{\uparrow\uparrow}(k)
\begin{pmatrix}
\psi_{\uparrow}(k) \\ \overline{\psi}_{\uparrow}(-k)
\end{pmatrix},
\end{equation}
with
\begin{equation}
\hat{M}_{\uparrow\uparrow}(k)=
\begin{pmatrix}
-i\omega_{n}+\xi_{\mib{k}} & \Delta_{\uparrow\uparrow}(k) \\
\overline{\Delta}_{\uparrow\uparrow}(k) & -i\omega_{n}-\xi_{-\mib{k}}
\end{pmatrix}.
\end{equation}
Integrating out the fermion fields in (\ref{freeenergy}), we obtain the explicit expression of the ``mean-field'' free-energy functional (assuming $\xi_{-\mib{k}}=\xi_{\mib{k}}$),
\begin{align}
&{\cal F}_{\rm MF}[\overline{\Delta},\Delta]=-\frac{1}{2\beta}\int_{r}\frac{\overline{\Delta}_{\uparrow\uparrow}(r)\Delta_{\uparrow\uparrow}(r)}{V(r)}
-\frac{1}{2\beta}\sum_{k}\ln\left[\frac{{\rm det}\,\hat{M}_{\uparrow\uparrow}(k)}{-(\omega_{n}^{2}+\xi_{\mib{k}}^{2})}\right],
\cr&\quad
{\rm det}\,\hat{M}_{\uparrow\uparrow}(k)=-[\omega_{n}^{2}+\xi_{\mib{k}}^{2}+\overline{\Delta}_{\uparrow\uparrow}(k)\Delta_{\uparrow\uparrow}(k)],
\label{expfreeenergy}
\end{align}
where we have measured the free energy from that in the normal state.
Note that since we have used no explicit relations for the odd-frequency pairing, the obtained free-energy functional is also valid for the even-frequency pairing as well.

\subsection{The correct choice of the saddle-point solution}

In the path integral, (\ref{pathaux}), the contours for $\overline{\Delta}_{\uparrow\uparrow}(1,2)$ and $\Delta_{\uparrow\uparrow}(1,2)$ can be taken independently\cite{Nagaosa99}.
However, in the saddle-point approximation, $\overline{\Delta}_{\uparrow\uparrow}(r)$ and $\Delta_{\uparrow\uparrow}(r)$ are no longer independent with each other.
Taking account of the fact that the free energy must be real, we have two possible choices of the extrema.
Namely, we have the choices\cite{deritwo},
\begin{equation}
\overline{\Delta}_{\uparrow\uparrow}(r)=+\Delta_{\uparrow\uparrow}(r)^{*},
\quad
\overline{\Delta}_{\uparrow\uparrow}(k)=+\Delta_{\uparrow\uparrow}(k)^{*},
\label{posrel}
\end{equation}
or
\begin{equation}
\overline{\Delta}_{\uparrow\uparrow}(r)=-\Delta_{\uparrow\uparrow}(r)^{*},
\quad
\overline{\Delta}_{\uparrow\uparrow}(k)=-\Delta_{\uparrow\uparrow}(k)^{*}.
\label{negrel}
\end{equation}
In order to demonstrate the difference between these two choices, let us consider a gap function of the form, $\Delta_{\uparrow\uparrow}(r)=\Delta_{0}\sin(\pi\tau/\beta)$ with a real $\Delta_{0}$ (an odd-frequency pairing) for example.
After some manipulation, we obtain the mean-field free energy near the transition temperature, $T_{c}$ as
\begin{equation}
{\cal F}_{\rm MF}=\pm a\left(\frac{T}{T_{c}}-1\right)\Delta_{0}^{2}+b\Delta_{0}^{4}+{\cal O}(\Delta_{0}^{6}),
\end{equation}
where $a$ and $b$ are positive constants.
The upper or the lower signs correspond to the choice (\ref{posrel}) or (\ref{negrel}).
From this consideration, we must choose the choice (\ref{posrel}) for a saddle-point solution, yielding $\Delta_{\uparrow\uparrow}(r)=\overline{\Delta}_{\uparrow\uparrow}(r)=0$ in the high-$T$ normal phase, while $\Delta_{\uparrow\uparrow}(r)$, $\overline{\Delta}_{\uparrow\uparrow}(r)\ne0$ in the low-$T$ superconducting phase.
Note that the relation (\ref{posrel}) is {\it implicitly} assumed in the work by Solenov {\it et al}. [eq.~(10) in ref.~\citen{Solenov09}].

Together with the condition (\ref{posrel}), the true saddle-point path is given by minimizing the free-energy functional (\ref{freeenergy}) as
\begin{equation}
0=\beta\frac{\delta {\cal F}_{\rm MF}}{\delta\overline{\Delta}_{\uparrow\uparrow}(r)}
=\Braket{\frac{\delta S_{\rm MF}}{\delta\overline{\Delta}_{\uparrow\uparrow}(r)}}_{\rm MF},
\quad
0=\Braket{\frac{\delta S_{\rm MF}}{\delta\Delta_{\uparrow\uparrow}(r)}}_{\rm MF},
\label{mfcondition}
\end{equation}
where we have defined the statistical average with respect to the mean-field,
\begin{equation}
\Braket{\cdots}_{\rm MF}=\frac{\int{\cal D}\overline{\psi}{\cal D}\psi\,e^{-S_{\rm MF}(\overline{\psi},\psi,\overline{\Delta},\Delta)}\cdots}{\int{\cal D}\overline{\psi}{\cal D}\psi\,e^{-S_{\rm MF}(\overline{\psi},\psi,\overline{\Delta},\Delta)}}.
\end{equation}
Equation~(\ref{mfcondition}) gives nothing but the gap equations,
\begin{equation}
\Delta_{\uparrow\uparrow}(r)=-V(r)F_{\uparrow\uparrow}(r),
\quad
\overline{\Delta}_{\uparrow\uparrow}(r)=-V(r)\overline{F}_{\uparrow\uparrow}(r),
\label{gapeq}
\end{equation}
where we have defined the anomalous green functions,
\begin{equation}
F_{\uparrow\uparrow}(r)=\Braket{\rho_{\uparrow\uparrow}(1,2)}_{\rm MF},
\quad
\overline{F}_{\uparrow\uparrow}(r)=\Braket{\overline{\rho}_{\uparrow\uparrow}(1,2)}_{\rm MF}.
\label{agreen}
\end{equation}

Once we choose the correct saddle point by (\ref{posrel}) and (\ref{gapeq}), it is easily confirmed with the help of the gap equations (\ref{gapeq}) that the associated anomalous green functions (\ref{agreen}) satisfy the relation,
\begin{equation}
\overline{F}_{\uparrow\uparrow}(r)=+F_{\uparrow\uparrow}(r)^{*},
\quad
\overline{F}_{\uparrow\uparrow}(k)=+F_{\uparrow\uparrow}(k)^{*},
\label{frel}
\end{equation}
where the fourier transforms of $F_{\uparrow\uparrow}(r)$ and $\overline{F}_{\uparrow\uparrow}(r)$ are given by
\begin{equation}
F_{\uparrow\uparrow}(k)=\Braket{\psi_{\uparrow}(k)\psi_{\uparrow}(-k)}_{\rm MF},
\quad
\overline{F}_{\uparrow\uparrow}(k)=\Braket{\overline{\psi}_{\uparrow}(-k)\overline{\psi}_{\uparrow}(k)}_{\rm MF}.
\end{equation}
We can also confirm this relation explicitly from the definition as was shown by Solenov and co-workers\cite{Solenov09}.

Now, let us make some comment on the relation between $\overline{F}_{\uparrow\uparrow}(k)$ and $F_{\uparrow\uparrow}(k)$.
For simplicity, we consider an on-site $s$-wave pairing since the frequency $i\omega_{n}$ dependence of the anomalous green functions is essential here.
If we were using the hamiltonian formalism, we could write down the spectral representation of the anomalous green functions as
\begin{align}
&\overline{F}_{\uparrow\uparrow}(i\omega_{n})=-\frac{1}{Z}\sum_{nm}\frac{e^{-\beta E_{m}}+e^{-\beta E_{n}}}{i\omega_{n}+E_{n}-E_{m}}\Braket{n|\psi_{\uparrow}^{\dagger}|m}\Braket{m|\psi_{\uparrow}^{\dagger}|n},
\cr&
F_{\uparrow\uparrow}(i\omega_{n})=-\frac{1}{Z}\sum_{nm}\frac{e^{-\beta E_{m}}+e^{-\beta E_{n}}}{i\omega_{n}+E_{n}-E_{m}}\Braket{n|\psi_{\uparrow}^{}|m}\Braket{m|\psi_{\uparrow}^{}|n},
\end{align}
where $Z=\sum_{n}e^{-\beta E_{n}}$.
From these expressions, we would immediately obtain the relation $\overline{F}_{\uparrow\uparrow}(i\omega_{n})=F_{\uparrow\uparrow}(-i\omega_{n})^{*}$.
Then, the property of the odd-frequency pairing would eventually lead to $\overline{F}_{\uparrow\uparrow}(i\omega_{n})=-F_{\uparrow\uparrow}(i\omega_{n})^{*}$, which seems to contradict with (\ref{frel}).
As was pointed out by Solenov {\it et al}.\cite{Solenov09}, however, there exists {\it no appropriate mean-field hamiltonian}, $H_{\rm MF}$ in the presence of the retardation of the gap function, although the above argument is based on the relation, $H_{\rm MF}\Ket{n}=E_{n}\Ket{n}$ and $\Bra{n}H_{\rm MF}=\Bra{n}E_{n}$.
Therefore, the above proof cannot work out due to the explicit use of the mean-field hamiltonian.
Note also that if we chose the maximum solution of the saddle-point (\ref{negrel}), we would obtain the incorrect relation, $\overline{F}_{\uparrow\uparrow}(i\omega_{n})=-F_{\uparrow\uparrow}(i\omega_{n})^{*}$.
It is the relation that has been widely believed to hold in the case of the odd-frequency pairing.
It should be also emphasized that even for the even-frequency pairing the ordinary proof for $\overline{F}_{\uparrow\uparrow}(i\omega_{n})=F_{\uparrow\uparrow}(-i\omega_{n})^{*}=F_{\uparrow\uparrow}(i\omega_{n})^{*}$ with the explicit use of $H_{\rm MF}$ should not work in the presence of the retardation.
Unfortunately, we did not realize the potential problem on this aspect since in the case of the even-frequency pairing the relation in question is continuously connected to the one with no retardation effect where the proof with $H_{\rm MF}$ should work out.
It only becomes evident when we have considered the odd-frequency pairing.

The discussions so far made in literatures (except ref.~\citen{Solenov09}) have begun with the incorrect relation, $\overline{F}_{\uparrow\uparrow}(i\omega_{n})=-F_{\uparrow\uparrow}(i\omega_{n})^{*}$, and the relation between $\overline{\Delta}_{\uparrow\uparrow}(k)$ and $\Delta_{\uparrow\uparrow}(k)$ are determined {\it afterward} so as to be consistent with the incorrect relation.
This is the source of various drawbacks of the odd-frequency pairing.
However, since the anomalous green function should be defined with respect to the given mean-field gap function with broken U(1) gauge symmetry, we must {\it first} set up the relation between $\overline{\Delta}_{\uparrow\uparrow}(k)$ and $\Delta_{\uparrow\uparrow}(k)$ by the minimum condition of the free energy as (\ref{posrel}), and the anomalous green functions are determined {\it afterward}.
With this procedure, we always obtain $\overline{\Delta}_{\uparrow\uparrow}(k)=\Delta_{\uparrow\uparrow}(k)^{*}$ and $\overline{F}_{\uparrow\uparrow}(k)=F_{\uparrow\uparrow}(k)^{*}$ both for the even-frequency and the odd-frequency pairing.
In summary, there are no formal differences in various formula between the even- and the odd-frequency pairings except the symmetry in the frequency dependences.

\subsection{Meissner kernel}

Let us consider the magnetic-field response.
For simplicity, we consider the London limit where a spatial dependence in $\mib{A}$ can be neglected.
With the London gauge $\mib{\nabla}\cdot\mib{A}=0$, the effect of the magnetic field is taken account by replacing the gauge-invariant minimal coupling as $\mib{\nabla}\to\mib{\nabla}+(ie/c)\mib{A}$ in $\xi$ ($e>0$).
Namely, we replace the matrix $\hat{M}_{\uparrow\uparrow}(k)$ with
\begin{equation}
\hat{M}_{\uparrow\uparrow}(k;\mib{A})=\hat{M}_{\uparrow\uparrow}(k)+\frac{e}{2mc^{2}}
\begin{pmatrix}
2c\mib{k}\cdot\mib{A}+e\mib{A}^{2} & 0 \\
0 & 2c\mib{k}\cdot\mib{A}-e\mib{A}^{2}
\end{pmatrix}.
\end{equation}
Then, the expansion of the free energy up to ${\cal O}(\mib{A}^{2})$ yields
\begin{equation}
{\cal F}_{\rm MF}(\mib{A})={\cal F}_{\rm MF}+\frac{1}{2}\frac{e^{2}}{mc^{2}}\sum_{ij}^{x,y,z}\tilde{K}_{ij}(0)A_{i}A_{j},
\end{equation}
where the Meissner kernel has been introduced by
\begin{equation}
\tilde{K}_{ij}(0)=\frac{1}{\beta}\sum_{k}
\left[
\frac{k_{i}k_{j}}{m}\left\{
G_{\uparrow\uparrow}(k)^{2}+F_{\uparrow\uparrow}(k)F_{\uparrow\uparrow}(k)^{*}
\right\}+\delta_{ij}G_{\uparrow\uparrow}(k)
\right].
\end{equation}
The current density is then obtained as
\begin{equation}
j_{i}=-c\frac{\partial{\cal F}_{\rm MF}(\mib{A})}{\partial A_{i}}=-\frac{e^{2}}{mc}\sum_{j}\tilde{K}_{ij}(0)A_{j}.
\end{equation}
Note that we have used the relations (\ref{posrel}) and (\ref{frel}).
Here, the explicit form of the green functions is given by
\begin{align}
&G_{\uparrow\uparrow}(k)=\frac{-(i\omega_{n}+\xi_{\mib{k}})}{\omega_{n}^{2}+\xi_{\mib{k}}^{2}+|\Delta_{\uparrow\uparrow}(k)|^{2}},
\cr
&F_{\uparrow\uparrow}(k)=\frac{\Delta_{\uparrow\uparrow}(k)}{\omega_{n}^{2}+\xi_{\mib{k}}^{2}+|\Delta_{\uparrow\uparrow}(k)|^{2}},
\end{align}
which are obtained by the matrix inverse as
\begin{equation}
\begin{pmatrix}
-G_{\uparrow\uparrow}(k) & F_{\uparrow\uparrow}(k) \\
F_{\uparrow\uparrow}(k)^{*} & G_{\uparrow\uparrow}(-k) \\
\end{pmatrix}
=\hat{M}_{\uparrow\uparrow}(k)^{-1}.
\end{equation}
Since the above expressions are exactly the same as the conventional ones\cite{AGD} even in the case of the odd-frequency pairing, the Meissner kernel (and the superfluid density) is positive as the consequence of the relation (\ref{posrel}).

\subsection{Analytic continuation to the real axis}

Owing to the general analytic property of the green functions, the gap function can be expressed in the form (omit the index $\mib{k}$ for simplicity),
\begin{equation}
\Delta_{\uparrow\uparrow}(z)=\int_{-\infty}^{\infty}\frac{d\omega}{\pi}\frac{\Delta_{\uparrow\uparrow}''(\omega)}{\omega-z},
\quad
\overline{\Delta}_{\uparrow\uparrow}(z)=\int_{-\infty}^{\infty}\frac{d\omega}{\pi}\frac{\overline{\Delta}_{\uparrow\uparrow}''(\omega)}{\omega-z},
\end{equation}
where $\Delta_{\uparrow\uparrow}''(\omega)={\rm Im}\,\Delta_{\uparrow\uparrow}(\omega+i0)$ and $\overline{\Delta}_{\uparrow\uparrow}''(\omega)={\rm Im}\,\overline{\Delta}_{\uparrow\uparrow}(\omega+i0)$.
By the relation $\Delta_{\uparrow\uparrow}(-i\omega_{n})=\phi \Delta_{\uparrow\uparrow}(i\omega_{n})$ ($\phi=+1$: the even-frequency pairing, $\phi=-1$: the odd-frequency pairing), we have
\begin{equation}
\Delta_{\uparrow\uparrow}''(-\omega)=-\phi\Delta_{\uparrow\uparrow}''(\omega),
\quad
\overline{\Delta}_{\uparrow\uparrow}''(-\omega)=-\phi\overline{\Delta}_{\uparrow\uparrow}''(\omega).
\end{equation}
From this relation, $\Delta_{\uparrow\uparrow}(i\omega_{n})$ and $\overline{\Delta}_{\uparrow\uparrow}(i\omega_{n})$ is real (pure imaginary) for the even-frequency (the odd-frequency) pairing.
It is also shown that the retarded functions satisfy the relation,
\begin{equation}
\Delta_{\uparrow\uparrow}(-\omega)=\phi\Delta_{\uparrow\uparrow}(\omega)^{*},
\quad
\overline{\Delta}_{\uparrow\uparrow}(-\omega)=\phi\overline{\Delta}_{\uparrow\uparrow}(\omega)^{*}.
\end{equation}
Thus, the terms ``even-frequency'' and ``odd-frequency'' represent the symmetry property of the Matsubara and the real part of the retarded gap functions.

Moreover, the minimum-energy condition (\ref{posrel}) demands
\begin{equation}
\overline{\Delta}_{\uparrow\uparrow}''(\omega)=\phi\Delta_{\uparrow\uparrow}''(\omega),
\end{equation}
and then the retarded functions satisfy
\begin{equation}
\overline{\Delta}_{\uparrow\uparrow}(\omega)=\phi\Delta_{\uparrow\uparrow}(\omega).
\end{equation}
Therefore, the analytic continuation from the upper half plane to the real axis is carried out by the replacement,
\begin{equation}
\Delta_{\uparrow\uparrow}(i\omega_{n})\to\Delta_{\uparrow\uparrow}(\omega),
\quad
\overline{\Delta}_{\uparrow\uparrow}(i\omega_{n})=\Delta_{\uparrow\uparrow}(i\omega_{n})^{*}\to\phi\Delta_{\uparrow\uparrow}(\omega).
\end{equation}
Then, the retarded green functions are given by
\begin{align}
&G_{\uparrow\uparrow}(\mib{k},\omega)=\frac{\omega+\xi_{\mib{k}}}{\omega^{2}-\xi_{\mib{k}}^{2}-\phi\Delta_{\uparrow\uparrow}(\mib{k},\omega)^{2}+i0{\rm sgn}(\omega)},
\cr
&F_{\uparrow\uparrow}(\mib{k},\omega)=\frac{-\Delta_{\uparrow\uparrow}(\mib{k},\omega)}{\omega^{2}-\xi_{\mib{k}}^{2}-\phi\Delta_{\uparrow\uparrow}(\mib{k},\omega)^{2}+i0{\rm sgn}(\omega)}.
\end{align}
Here the global phase of the gap function due to broken U(1) gauge symmetry does not appear in the denominator of the green functions.
Note that the retarded anomalous green functions satisfy the similar relations of the gap functions, and $\phi\Delta(\mib{k},\omega=0)^{2}\ge0$.

\section{Summary}
We have derived the free energy of the superconducting state as a saddle-point solution in the path-integral framework.
The emphasis has been placed on the correct choice of the minimum free-energy solution that provides us with the ordinary relation between the gap function and its conjugate counterpart,
\[
\Delta^{\dagger}_{\beta\alpha}(k)\equiv\overline{\Delta}_{\alpha\beta}(k)=\Delta_{\alpha\beta}(k)^{*}.
\]
The associated anomalous green functions determined in a consistent way through the gap equations satisfy the similar relation,
\[
F^{+}_{\beta\alpha}(k)\equiv\overline{F}_{\alpha\beta}(k)=F_{\alpha\beta}(k)^{*},
\]
irrespective of the symmetry of the gap functions in the frequency domain.
A naive derivation for this relation using the mean-field hamiltonian should not be valid in the presence of the retardation in the gap functions.

The gap function in Matsubara representation is pure imaginary for the odd-frequency pairing.
The analytic continuation from the upper half plane to the real axis is carried out by the replacement,
\begin{align}
&\Delta_{\alpha\beta}(\mib{k},i\omega_{n})\to\Delta_{\alpha\beta}(\mib{k},\omega),
\cr
&\Delta^{\dagger}_{\beta\alpha}(\mib{k},i\omega_{n})=\Delta_{\alpha\beta}(\mib{k},i\omega_{n})^{*}\to-\Delta_{\alpha\beta}(\mib{k},\omega),
\end{align}
with $\Delta_{\alpha\beta}(\mib{k},-\omega)=-\Delta_{\alpha\beta}(\mib{k},\omega)^{*}$ for the odd-frequency pairing.

From now on, the odd-frequency pairing has no apparent deficiency, and hence a proper strongly retarded interaction can mediate the odd-frequency pairing in real materials.

\section*{Acknowledgments}
This work was supported by a Grant-in-Aid for Scientific Research on Innovative Areas ``Heavy Electrons" (No.20102008) of The Ministry of Education, Culture, Sports, Science, and Technology (MEXT), Japan.
One of the authors (K.M.) is supported in part by a Grant-in-Aid for Scientific Research on Innovative Areas ``Topological Quantum Phenomena" (No.22103003) of MEXT, and by a Grant-in-Aid for Scientific Research (No.19340099) of the Japan Society for the Promotion of Science (JSPS).
Y.F. is supported by a Grant-in-Aid for Young Scientists (No.21840035) of JSPS.

\appendix
\section{In the case of general type of pairings}
In this appendix, we repeat the discussions in the case of general type of pairings.
The formula obtained here are valid both for the even- and the odd-frequency pairings with coexistence of the singlet and the triplet states in the case of inhomogeneity.

We consider the general form of the irreducible interaction in the path integral (\ref{pathz}), \begin{align}
&S_{0}=\int_{1}\overline{\psi}_{\alpha}(1)\left[\partial_{\tau}+\xi\right]\psi_{\alpha}(1),
\cr
\quad
&S_{\rm int}=\frac{1}{2\beta}\int_{1}\int_{2}V_{\alpha\beta;\gamma\delta}(1-2)\overline{\rho}_{\alpha\beta}(1,2)\rho_{\gamma\delta}(1,2),
\end{align}
where the summation is understood for the repeated spin indices (the greek letters).
Due to the hermiticity, the pairing interaction satisfies
\begin{equation}
V_{\alpha\beta;\gamma\delta}(1-2)=V_{\gamma\delta;\alpha\beta}(1-2)^{*}.
\end{equation}
Here, the pair field and its complex conjugate have been defined by
\begin{align}
&\rho_{\alpha\beta}(1,2)=\psi_{\alpha}(1)\psi_{\beta}(2),
\cr
&\overline{\rho}_{\alpha\beta}(1,2)=\overline{\psi}_{\beta}(2)\overline{\psi}_{\alpha}(1).
\end{align}
The anti-commutation relation (Pauli principle) is expressed as
\begin{equation}
\rho_{\alpha\beta}(1,2)=-\rho_{\beta\alpha}(2,1),
\quad
\overline{\rho}_{\alpha\beta}(1,2)=-\overline{\rho}_{\beta\alpha}(2,1).
\label{paulipr}
\end{equation}

By the Stratonovich-Hubbard transformation, we obtain
\begin{align}
&S_{\rm aux}
=-\frac{1}{2\beta}\int_{1}\int_{2}\left[
\overline{\rho}_{\alpha\beta}(1,2)\Delta_{\alpha\beta}(1,2)
+\overline{\Delta}_{\alpha\beta}(1,2)\rho_{\alpha\beta}(1,2)
\right],
\cr
&S_{\Delta}
=-\frac{1}{2\beta}\int_{1}\int_{2}\left[V(1-2)^{-1}\right]_{\alpha\beta;\gamma\delta}
\overline{\Delta}_{\alpha\beta}(1,2)\Delta_{\gamma\delta}(1,2),
\end{align}
where $V(1-2)^{-1}$ is the matrix inverse of $V(1-2)$ whose $(\alpha\beta)$-$(\gamma\delta)$ component is given by $V_{\alpha\beta;\gamma\delta}(1-2)$.
Note that the complex $c$-number auxiliary fields $\Delta_{\alpha\beta}(1,2)$ and $\overline{\Delta}_{\alpha\beta}(1,2)$ satisfy the Pauli principle as similar to (\ref{paulipr}).

As was explained in the main text, the correct minimum of the free energy within the saddle-point approximation is obtained by the gap functions that satisfy the gap equations,
\begin{align}
&\Delta_{\alpha\beta}(1,2)=-V_{\alpha\beta;\gamma\delta}(1-2)F_{\gamma\delta}(1,2),
\cr
&\overline{\Delta}_{\alpha\beta}(1,2)=-V_{\alpha\beta;\gamma\delta}(1-2)^{*}\overline{F}_{\gamma\delta}(1,2),
\end{align}
and the minimum-energy condition,
\begin{equation}
\Delta_{\alpha\beta}(1,2)^{*}=\overline{\Delta}_{\alpha\beta}(1,2)\equiv\Delta^{\dagger}_{\beta\alpha}(2,1).
\label{gapcond}
\end{equation}
Here we have introduced the anomalous green functions defined by
\begin{align}
&F_{\alpha\beta}(1,2)=\Braket{\psi_{\alpha}(1)\psi_{\beta}(2)}_{\rm MF},
\cr
&\overline{F}_{\alpha\beta}(1,2)=\Braket{\overline{\psi}_{\beta}(2)\overline{\psi}_{\alpha}(1)}_{\rm MF}\equiv F^{+}_{\beta\alpha}(2,1),
\label{agf}
\end{align}
and $\Delta^{\dagger}$ and $F^{+}$ are used for notational simplicity.
From the minimum-energy condition (\ref{gapcond}) with the help of the gap equations, we obtain the relation,
\begin{equation}
F_{\alpha\beta}(1,2)^{*}=\overline{F}_{\alpha\beta}(1,2)=F_{\beta\alpha}^{+}(2,1).
\label{fcond}
\end{equation}

Integrating out the fermion fields by noting the quadratic form of the action,
\begin{align}
&S_{0}+S_{\rm aux}=\frac{1}{2\beta^{2}}\int_{1}\int_{2}
\begin{pmatrix}
\overline{\psi}_{\alpha}(1), & \psi_{\alpha}(1)
\end{pmatrix}
\hat{M}_{\alpha\beta}(1,2)
\begin{pmatrix}
\psi_{\beta}(2) \\ \overline{\psi}_{\beta}(2)
\end{pmatrix},
\cr&\quad
\hat{M}_{\alpha\beta}(1,2)=
\begin{pmatrix}
\beta^{2}\delta(1-2)\delta_{\alpha\beta}(\partial_{\tau}+\xi) & \beta\Delta_{\alpha\beta}(1,2) \\
\beta\Delta_{\alpha\beta}^{\dagger}(1,2) & \beta^{2}\delta(1-2)\delta_{\alpha\beta}(\partial_{\tau}-\xi)
\end{pmatrix},
\label{mmatgen}
\end{align}
we obtain the free energy as
\begin{multline}
{\cal F}_{\rm MF}=-\frac{1}{2\beta^{2}}\int_{1}\int_{2}\left[V(1-2)^{-1}\right]_{\alpha\beta;\gamma\delta}\overline{\Delta}_{\alpha\beta}(1,2)\Delta_{\gamma\delta}(1,2)
\\
-\frac{1}{2\beta}\ln\left(
\frac{{\rm det}\hat{M}}{{\rm det}\hat{M}_{0}}
\right).
\label{fegen}
\end{multline}
where $\hat{M}_{0}\equiv\hat{M}(\Delta=\Delta^{\dagger}=0)$.

Let us introduce the green functions together with (\ref{agf}),
\begin{align}
&G_{\alpha\beta}(1,2)=-\Braket{\psi_{\alpha}(1)\overline{\psi}_{\beta}(2)}_{\rm MF},
\cr
&G'_{\alpha\beta}(1,2)=-\Braket{\overline{\psi}_{\alpha}(1)\psi_{\beta}(2)}_{\rm MF}
=-G_{\beta\alpha}(2,1).
\end{align}
When the action is quadratic, the matrix $\hat{M}$ and the green functions are related as
\begin{equation}
\hat{M}_{\alpha\beta}(1,2)=
\begin{pmatrix}
-G_{\alpha\beta}(1,2) & F_{\alpha\beta}(1,2) \\
F^{+}_{\alpha\beta}(1,2) & -G_{\alpha\beta}'(1,2)
\end{pmatrix}^{-1},
\label{gmat}
\end{equation}
and it gives the Gor'kov equations,
\begin{align}
&G_{\alpha\beta}(1,2)=G^{(0)}_{\alpha\beta}(1,2)-\frac{1}{\beta}\int_{3}\int_{4}G^{(0)}_{\alpha\gamma}(1,3)\Delta_{\gamma\delta}(3,4)F^{+}_{\delta\beta}(4,2),
\cr
&G'_{\alpha\beta}(1,2)=G^{(0)'}_{\alpha\beta}(1,2)-\frac{1}{\beta}\int_{3}\int_{4}G^{(0)'}_{\alpha\gamma}(1,3)\Delta^{\dagger}_{\gamma\delta}(3,4)F_{\delta\beta}(4,2),
\cr
&F_{\alpha\beta}(1,2)=-\frac{1}{\beta}\int_{3}\int_{4}G^{(0)}_{\alpha\gamma}(1,3)\Delta_{\gamma\delta}(3,4)G'_{\delta\beta}(4,2),
\cr
&F^{+}_{\alpha\beta}(1,2)=-\frac{1}{\beta}\int_{3}\int_{4}G^{(0)'}_{\alpha\gamma}(1,3)\Delta^{\dagger}_{\gamma\delta}(3,4)G_{\delta\beta}(4,2),
\end{align}
where we have introduced the bare green functions,
\begin{align}
&G^{(0)}_{\alpha\beta}(1,2)^{-1}=-\beta^{2}\delta(1-2)\delta_{\alpha\beta}(\partial_{\tau}+\xi),
\cr
&G^{(0)'}_{\alpha\beta}(1,2)^{-1}=-\beta^{2}\delta(1-2)\delta_{\alpha\beta}(\partial_{\tau}-\xi).
\end{align}

Now, let us introduce the fourier transformation according to the first line of (\ref{ftsimple}), then we obtain
\begin{align}
&G_{\alpha\beta}(k,k')=-\Braket{\psi_{\alpha}(k)\overline{\psi}_{\beta}(k')}_{\rm MF},
\cr
&G'_{\alpha\beta}(k,k')=-\Braket{\overline{\psi}_{\alpha}(-k)\psi_{\beta}(-k')}_{\rm MF},
\cr
&F_{\alpha\beta}(k,k')=\Braket{\psi_{\alpha}(k)\psi_{\beta}(-k')}_{\rm MF},
\cr
&F^{+}_{\alpha\beta}(k,k')=\Braket{\overline{\psi}_{\alpha}(-k)\overline{\psi}_{\beta}(k')}_{\rm MF}=\overline{F}_{\beta\alpha}(k',k).
\end{align}
Here, $A(k,k')=\beta^{-1}\int_{1}\int_{2}A(1,2)e^{-i(kx_{1}-k'x_{2})}$.
The bare green functions are given by
\begin{align}
&G_{\alpha\beta}^{(0)}(k,k')=(i\omega_{n}-\xi_{\mib{k}})^{-1}\delta_{kk'}\delta_{\alpha\beta}
=-G_{\beta\alpha}^{(0)'}(-k',-k).
\end{align}
For the gap functions, we define the fourier transformation as 
\begin{align}
&\Delta_{\alpha\beta}(1,2)=\sum_{kk'}\Delta_{\alpha\beta}(k,k')e^{i(kx_{1}-k'x_{2})},
\cr
&\overline{\Delta}_{\alpha\beta}(1,2)=\sum_{kk'}\overline{\Delta}_{\alpha\beta}(k,k')e^{-i(kx_{1}-k'x_{2})},
\cr
&\Delta^{\dagger}_{\alpha\beta}(1,2)=\sum_{kk'}\Delta^{\dagger}_{\alpha\beta}(k,k')e^{i(kx_{1}-k'x_{2})}.
\end{align}
The minimum-energy condition (\ref{gapcond}) and (\ref{fcond}) become
\begin{align}
&\Delta_{\alpha\beta}(k,k')^{*}=\overline{\Delta}_{\alpha\beta}(k,k')=\Delta_{\beta\alpha}^{\dagger}(k',k),
\cr
&F_{\alpha\beta}(k,k')^{*}=\overline{F}_{\alpha\beta}(k,k')=F_{\beta\alpha}^{+}(k',k).
\end{align}

In the case of a homogenous state, we can set $F_{\alpha\beta}(k,k')=F_{\alpha\beta}(k)\delta_{kk'}$, $\Delta_{\alpha\beta}(k,k')=\Delta_{\alpha\beta}(k)\delta_{kk'}$ and so on.
Using the momentum-frequency basis, the free energy (\ref{fegen}) becomes
\begin{multline}
{\cal F}_{\rm MF}=-\frac{1}{2}\sum_{kk'}W_{\alpha\beta;\gamma\delta}(k-k')\Delta_{\alpha\beta}(k)^{*}\Delta_{\gamma\delta}(k')
\\
-\frac{1}{2\beta}\sum_{k}\ln\left(
\frac{{\rm det}\,\hat{M}(k)}{{\rm det}\,\hat{M}_{0}(k)}
\right),
\end{multline}
where $[V(r)^{-1}]_{\alpha\beta;\gamma\delta}=\sum_{q}W_{\alpha\beta;\gamma\delta}(q)e^{iqr}$.
Using the gap equation,
\begin{equation}
\Delta_{\alpha\beta}(k)=-\frac{1}{\beta}\sum_{k'}V_{\alpha\beta;\gamma\delta}(k-k')F_{\gamma\delta}(k'),
\end{equation}
the free energy is rewritten as
\begin{equation}
{\cal F}_{\rm MF}=\frac{1}{2\beta}\sum_{k}\left[
\Delta_{\alpha\beta}(k)^{*}F_{\alpha\beta}(k)-\ln\left(
\frac{{\rm det}\,\hat{M}(k)}{{\rm det}\,\hat{M}_{0}(k)}
\right)
\right].
\end{equation}

When we decompose the gap function into the singlet component $d_{0}(k)$ and the triplet component $\mib{d}(k)$ as
\begin{equation}
\Delta_{\alpha\beta}(k)=d_{0}(k)(i\sigma^{y})_{\alpha\beta}+\mib{d}(k)\cdot(i\mib{\sigma}\sigma^{y})_{\alpha\beta},
\end{equation}
the explicit expressions are obtained as follows:
\begin{equation}
{\rm det}\,\hat{M}(k)=[\omega_{n}^{2}+E_{+}(k)^{2}][\omega_{n}^{2}+E_{-}(k)^{2}],
\end{equation}
with
\begin{align}
&D_{0}(k)=|d_{0}(k)|^{2}+|\mib{d}(k)|^{2},
\cr
&\mib{D}(k)=d_{0}(k)\mib{d}(k)^{*}+\mib{d}(k)d_{0}(k)^{*}+i\left[\mib{d}(k)\times\mib{d}(k)^{*}\right],
\cr
&E_{\pm}(k)=\sqrt{\xi_{\mib{k}}^{2}+D_{0}(k)\pm|\mib{D}(k)|}.
\end{align}
Note that $D_{0}(k)$ and $\mib{D}(k)$ are real functions, and the Pauli principle demands,
\begin{align}
&d_{0}(k)=d_{0}(-k)=\pm d_{0}(-\mib{k},i\omega_{n}),
\cr
&\mib{d}(k)=-\mib{d}(-k)=\mp\mib{d}(-\mib{k},i\omega_{n}),
\end{align}
where the upper and the lower signs correspond to the even-frequency and the odd-frequency pairings, respectively.
In the case of the unitary pairing, i.e., $\mib{D}(k)=0$, no time-reversal symmetry breaking occurs even for the odd-frequency pairing (it breaks only the {\it imaginary}-time-reversal symmetry of the relative coordinate).

The green functions are written as
\begin{align}
&G_{\alpha\beta}(k)=-G'_{\beta\alpha}(-k)=G_{0}(k)\delta_{\alpha\beta}+\mib{G}(k)\cdot\mib{\sigma}_{\alpha\beta},
\cr
&F_{\alpha\beta}(k)=F^{+}_{\beta\alpha}(k)^{*}=F_{0}(k)(i\sigma^{y})_{\alpha\beta}+\mib{F}(k)\cdot(i\mib{\sigma}\sigma^{y})_{\alpha\beta},
\cr&
\end{align}
where the scalar and the vector components are given by
\begin{align}
&G_{0}(k)=-\frac{(i\omega_{n}+\xi_{\mib{k}})[\omega_{n}^{2}+\xi_{\mib{k}}^{2}+D_{0}(k)]}{[\omega_{n}^{2}+E_{+}(k)^{2}][\omega_{n}^{2}+E_{-}(k)^{2}]},
\cr
&\mib{G}(k)=\frac{(i\omega_{n}+\xi_{\mib{k}})\mib{D}(k)}{[\omega_{n}^{2}+E_{+}(k)^{2}][\omega_{n}^{2}+E_{-}(k)^{2}]},
\cr
&F_{0}(k)=\frac{(\omega_{n}^{2}+\xi_{\mib{k}}^{2})d_{0}(k)+[d_{0}(k)^{2}-\mib{d}(k)^{2}]d_{0}(k)^{*}}{[\omega_{n}^{2}+E_{+}(k)^{2}][\omega_{n}^{2}+E_{-}(k)^{2}]},
\cr
&\mib{F}(k)=\frac{(\omega_{n}^{2}+\xi_{\mib{k}}^{2})\mib{d}(k)-[d_{0}(k)^{2}-\mib{d}(k)^{2}]\mib{d}(k)^{*}}{[\omega_{n}^{2}+E_{+}(k)^{2}][\omega_{n}^{2}+E_{-}(k)^{2}]}.
\end{align}

For the magnetic-field response, we again introduce the minimal coupling in $\xi$ of (\ref{mmatgen}), and expand the free energy up to ${\cal O}(\mib{A}^{2})$ by using (\ref{gmat}).
Then, we obtain
\begin{equation}
{\cal F}_{\rm MF}(\mib{A})={\cal F}_{\rm MF}+\frac{1}{2}\frac{e^{2}}{mc^{2}}\sum_{ij}^{x,y,z}\sum_{q}\tilde{K}_{ij}(q)A_{i}(q)A_{j}(-q),
\end{equation}
where the Meissner kernel has been introduced by
\begin{multline}
\tilde{K}_{ij}(q)=\frac{1}{\beta}\sum_{k}
\biggl[
\frac{k_{i}k_{j}}{m}\left\{
G_{\alpha\beta}(k)G_{\beta\alpha}(k-q)+F^{}_{\alpha\beta}(k)F^{+}_{\beta\alpha}(k-q)
\right\}
\\
+\delta_{ij}G_{\alpha\alpha}(k)
\biggr],
\end{multline}
and $q=(\mib{q},i\epsilon_{m})$ with the bosonic Matsubara frequency, $\epsilon_{m}=2m\pi/\beta$.


\begin{thebibliography}{99}
\bibitem{inhomo} Cases concerning the odd-frequency pair potential are different issues from those given in this paper.
\bibitem{deritwo} To see this for example, we focus on the first term of (\ref{expfreeenergy}). Putting $\overline{\Delta}_{\uparrow\uparrow}(r)=\Delta_{\uparrow\uparrow}(r)^{*}e^{i\varphi(r)}$, we immediately have two choices $\varphi(r)=0,\pi$.
\end{thebibliography}
\end{document}